\documentclass[compsoc,conference,a4paper,10pt,times,twoside]{IEEEtran}
\usepackage{cite}
\usepackage{amsmath,amssymb,amsfonts}
\usepackage{graphicx}
\usepackage{caption}

\newcommand{\CSnHFILThreads}{{67}}
\newcommand{\CSnHFPSThreads}{{33}}
\newcommand{\CSnHFILPSOverlapsThreads}{{23}}

\newcommand{\CSnHFILPSFinalThreads}{{80}}
\newcommand{\CSnILPSHighlyRelevantThreads}{{16}}
\newcommand{\CSnILPSHighlyRelevantPosts}{{742}}
\newcommand{\CSnILPSLowerRelevantThreads}{{64}}
\newcommand{\CSnILPSLowerRelevantPosts}{{173}}
\newcommand{\CSnILPSAllRelevantPosts}{{915}}
\newcommand{\CSnILPSAllRelevantUsers}{{187}}

\newcommand{\CSnILPSRelatedPostsSignificanceTestMethod}{{Kruskal-Wallis}}
\newcommand{\CSnILPSRelatedPostsSignificanceTestDoF}{{2}}
\newcommand{\CSnILPSRelatedPostsSignificanceTestStat}{{80.32}}
\newcommand{\CSnILPSRelatedPostsSignificanceTestPValue}{{p < .0001}}

\newcommand{\CSnILPSRelatedPostsEffectSize}{{0.43}}

\newcommand{\CSnILPSRelatedPostsPostHocPValueONETHREE}{{p < .001}}

\newcommand{\CSnILPSRelatedPostsPostHocPValueTWOTHREE}{{p < .0001}}

\newcommand{\CSnILPSRelatedUsersSignificanceTestMethod}{{Kruskal-Wallis}}
\newcommand{\CSnILPSRelatedUsersSignificanceTestDoF}{{2}}
\newcommand{\CSnILPSRelatedUsersSignificanceTestStat}{{78.07}}
\newcommand{\CSnILPSRelatedUsersSignificanceTestPValue}{{p < .0001}}

\newcommand{\CSnILPSRelatedUsersEffectSize}{{0.42}}

\newcommand{\CSnILPSRelatedUsersPostHocPValueONETHREE}{{p < .001}}

\newcommand{\CSnILPSRelatedUsersPostHocPValueTWOTHREE}{{p < .0001}}

\newcommand{\CSnHFILPSNPostsTopCategoriesProportion}{{96.39}}

\newcommand{\CSnHFILPSNPostsTopCategoriesNumberOneProportion}{{68.63}}

\newcommand{\CSnHFILPSNPostsTopCategoriesNumberTwoProportion}{{12.90}}

\newcommand{\CSnewilpsnIsraelWarDefacementFunMotives}{{221}}
\newcommand{\CSnewilpsnIsraelWarDefacementFunMotivesProps}{{12.17}}

\newcommand{\CSnewilpsnIsraelWarDefacementNationalisticConflictsMotives}{{three}}

\newcommand{\CSnewilpsnIsraelWarDefacementProILMotives}{{one}}

\newcommand{\CSnewilpsnIsraelWarDefacementProPSMotives}{{559}}
\newcommand{\CSnewilpsnIsraelWarDefacementProPSMotivesProps}{{30.78}}

\newcommand{\CSnewilpsnIsraelWarDefacementFinanceMotives}{{two}}

\newcommand{\CSnewilpsnIsraelWarTopOneDefacerProps}{{16.46}}
\newcommand{\CSnewilpsnIsraelWarTopTenDefacerProps}{{55.67}}
\newcommand{\CSnewilpsnIsraelWarDefacementsTotalAttacksHittingILandPS}{{1\,816}}
\newcommand{\CSnewilpsnIsraelWarTotalDefacements}{{105\,432}}

\newcommand{\CSnewilpsnIsraelWarDomainCatCOIL}{{1\,384}}
\newcommand{\CSnewilpsnIsraelWarDomainCatCOILProps}{{76.21}}
\newcommand{\CSnewilpsnIsraelWarDomainCatORGIL}{{61}}
\newcommand{\CSnewilpsnIsraelWarDomainCatORGILProps}{{3.36}}

\newcommand{\CSnewilpsnIsraelWarDomainCatACIL}{{16}}
\newcommand{\CSnewilpsnIsraelWarDomainCatACILProps}{{0.88}}
\newcommand{\CSnewilpsnIsraelWarDomainCatCOM}{{262}}

\newcommand{\CSnewilpsnIsraelWarDomainCatORG}{{11}}

\newcommand{\CSnewilpsnIsraelWarDomainCatNET}{{22}}

\newcommand{\CSnewilpsnIsraelWarDomainCatBIZ}{{4}}

\newcommand{\CSnewilpsnIsraelWarDomainCatPS}{{25}}

\newcommand{\CSnewilpsnIsraelWarDomainCatCLUB}{{3}}

\newcommand{\CSnewilpsnIsraelWarDomainCatRemaining}{{22}}

\newcommand{\CSnewilpsnIsraelWarDefacementsHittingIsrael}{{1\,791}}

\newcommand{\CSnewilpsnIsraelWarDefacersHittingIsrael}{{439}}

\newcommand{\CSnewilpsnIsraelWarDefacementsHittingPalestine}{{25}}

\newcommand{\CSnewilpsnIsraelWarDefacersHittingPalestine}{{18}}

\newcommand{\CSnewilpsnIsraelWarDomainCatART}{{1}}

\newcommand{\CSnewilpsnDefacementsILSignificanceTestMethod}{{Kruskal-Wallis}}
\newcommand{\CSnewilpsnDefacementsILSignificanceTestDoF}{{2}}
\newcommand{\CSnewilpsnDefacementsILSignificanceTestStat}{{62.33}}
\newcommand{\CSnewilpsnDefacementsILSignificanceTestPValue}{{p < .0001}}

\newcommand{\CSnewilpsnDefacementsILEffectSize}{{0.33}}

\newcommand{\CSnewilpsnDefacementsILPostHocPValueONETWO}{{p < .0001}}

\newcommand{\CSnewilpsnDefacementsILPostHocPValueONETHREE}{{p < .0001}}

\newcommand{\CSnewilpsnDefacementsILPostHocPValueTWOTHREE}{{p < .0001}}

\newcommand{\CSnewilpsnDefacementsPSSignificanceTestDoF}{{2}}
\newcommand{\CSnewilpsnDefacementsPSSignificanceTestStat}{{10.50}}
\newcommand{\CSnewilpsnDefacementsPSSignificanceTestPValue}{{p < .01}}

\newcommand{\CSnewilpsnDefacementsPSEffectSize}{{0.05}}

\newcommand{\CSnewilpsnDefacementsPSPostHocPValueONETWO}{{p = .2007}}

\newcommand{\CSnewilpsnDefacementsPSPostHocPValueONETHREE}{{p < .05}}

\newcommand{\CSnewilpsnDefacementsPSPostHocPValueTWOTHREE}{{p < .01}}

\newcommand{\CSnewilpsnDDoSAttacksILSignificanceTestMethod}{{Kruskal-Wallis}}
\newcommand{\CSnewilpsnDDoSAttacksILSignificanceTestDoF}{{2}}
\newcommand{\CSnewilpsnDDoSAttacksILSignificanceTestStat}{{24.48}}
\newcommand{\CSnewilpsnDDoSAttacksILSignificanceTestPValue}{{p < .0001}}

\newcommand{\CSnewilpsnDDoSAttacksILEffectSize}{{0.12}}

\newcommand{\CSnewilpsnDDoSAttacksILPostHocPValueONETWO}{{p < .0001}}

\newcommand{\CSnewilpsnDDoSAttacksILPostHocPValueONETHREE}{{p < .01}}

\newcommand{\CSnewilpsnDDoSAttacksILPostHocPValueTWOTHREE}{{p < .01}}

\newcommand{\CSnewilpsnDDoSAttacksPSSignificanceTestDoF}{{2}}
\newcommand{\CSnewilpsnDDoSAttacksPSSignificanceTestStat}{{32.91}}
\newcommand{\CSnewilpsnDDoSAttacksPSSignificanceTestPValue}{{p < .0001}}

\newcommand{\CSnewilpsnDDoSAttacksPSEffectSize}{{0.17}}

\newcommand{\CSnewilpsnDDoSAttacksPSPostHocPValueONETWO}{{p < .0001}}

\newcommand{\CSnewilpsnDDoSAttacksPSPostHocPValueONETHREE}{{p < .001}}

\newcommand{\CSnewilpsnDDoSAttacksPSPostHocPValueTWOTHREE}{{p < .01}}

\usepackage{contour,ulem}

\contourlength{0.8pt}

\usepackage{tikz}

\usepackage{algorithm,algpseudocode,algorithmicx}
\algnewcommand\algorithmicforeach{\textbf{for each}}
\algdef{S}[FOR]{ForEach}[1]{\algorithmicforeach\ #1\ \algorithmicdo}
\algdef{SE}[DOWHILE]{Do}{doWhile}{\algorithmicdo}[1]{\algorithmicwhile\ #1}

\usepackage{booktabs} 
\usepackage{multirow,multicol} 
\usepackage{makecell} 
\usepackage[switch]{lineno} 
\usepackage{threeparttable} 
\usepackage{url}  
\usepackage{xspace} 
\usepackage{colortbl} 
\usepackage{fancyhdr} 
\DeclareUrlCommand\myurl{\urlstyle{tt}} 
\usepackage[hidelinks,unicode,linkcolor=black]{hyperref} 
\newcommand*{\doi}[1]{\href{https://doi.org/#1}{\myurl{DOI: #1}}}

\newcommand{\para}[1]{\vspace{0.75mm}\noindent\textbf{#1.}}
\newcommand{\lpara}[1]{\vspace{0.85mm}\noindent\textbf{#1.}}

\usepackage{adjustbox} 
\newcolumntype{R}[2]{>{\adjustbox{angle=#1,lap=\width-(#2)}\bgroup}l<{\egroup}}

\newcommand{\hackforums}{{\small\scshape Hack Forums}\xspace}

\newcommand{\zoneh}{{\small\scshape Zone-H}\xspace}

\newcommand{\zonexsec}{{\small\scshape Zone-Xsec}\xspace}

\newcommand{\haxorid}{{\small\scshape Haxor-ID}\xspace}

\newcommand{\defacerpro}{{\small\scshape Defacer-Pro}\xspace}

\newcommand{\ownzyou}{{\small\scshape OwnzYou}\xspace}

\newcommand{\crimebb}{{\small\scshape CrimeBB}\xspace}

\newcommand{\ccc}{Cambridge Cybercrime Centre\xspace}

\IEEEoverridecommandlockouts
\def\BibTeX{{\rm B\kern-.05em{\sc i\kern-.025em b}\kern-.08em
    T\kern-.1667em\lower.7ex\hbox{E}\kern-.125emX}}
\begin{document}

\pagestyle{fancy}
\fancyhead[RO,LE]{\small{In Proceedings of the IEEE European Symposium on Security and Privacy Workshops 2025\vspace{0.05mm}}}
\fancyfoot[LO,RE]{\small{Anh V. Vu, Alice Hutchings, and Ross Anderson}}
\fancyfoot[RO,LE]{\small{\thepage}}
\fancyfoot[CO,CE]{}

\title{Yet Another Diminishing Spark: Low-level Cyberattacks\\in the Israel-Gaza Conflict}
\author{\IEEEauthorblockN{Anh V. Vu}
\IEEEauthorblockA{University of Cambridge\\Cambridge Cybercrime Centre\\anh.vu@cl.cam.ac.uk}
\and
\IEEEauthorblockN{Alice Hutchings}
\IEEEauthorblockA{University of Cambridge\\Cambridge Cybercrime Centre\\alice.hutchings@cl.cam.ac.uk}
\and
\IEEEauthorblockN{Ross Anderson}
\IEEEauthorblockA{University of Cambridge\\and University of Edinburgh\\ross.anderson@cl.cam.ac.uk}}

\maketitle

\begin{abstract}
We report empirical evidence of web defacement and DDoS attacks carried out by low-level cybercrime actors in the Israel-Gaza conflict. Our quantitative measurements indicate an immediate increase in such cyberattacks following the Hamas-led assault and the subsequent declaration of war. However, the surges waned quickly after a few weeks, with patterns resembling those observed in the aftermath of the Russian invasion of Ukraine. The scale of attacks and discussions within the hacking community this time was both significantly lower than those during the early days of the Russia-Ukraine war, and attacks have been prominently one-sided: many pro-Palestinian supporters have targeted Israel, while attacks on Palestine have been much less significant. Beyond targeting these two, attackers also defaced sites of other countries to express their war support. Their broader opinions are also largely disparate, with far more support for Palestine and many objections expressed toward Israel.
\end{abstract}

\section{Introduction} \label{sec:introduction}
\noindent Cyber operations have been used in armed conflicts as part of `hybrid' warfare~\cite{hoffman2007conflict}, involving both high and low-level actors who have contributed to a digital confrontation in various ways. Since the Hamas attack on 7 October 2023 and the declaration of war, Israeli and Palestinian digital assets have been targeted, including distributed denial-of-service (DDoS) and website defacement attacks. For example, after the Jerusalem Post suffered a DDoS attack, geo-fencing measures were implemented to restrict access to users within Israel~\cite{jerusalempostddos}. Media outlets in Israel and the Palestinian territories have also been targeted~\cite{oxford2024cyberattacks}, with hundreds of journalists and media workers (unintentionally) killed~\cite{israelgazajournalistcasualties}. This ongoing war introduces an opportunity to analyse cyberattack patterns and assess how they differ from other conflicts such as the invasion of Ukraine in 2022, given that both are major wars rooted in national aspirations but with different political contexts.

While nation-state attacks are widely highlighted in the news, the role of low-level cybercrime actors and volunteer hacktivists appears to have been under-reported. Our prior work~\cite{vu2024getting} analysed `nationalistic' activities in the Russia-Ukraine war that began in February 2022, revealing swift responses with immediate but short-lived peaks of online discussions, website defacements, and DDoS attacks targeting both countries. We take a similar approach to analyse the non-governmental low-level cyberattacks in the Israel-Gaza conflict, focusing particularly on web defacement and UDP amplification DDoS attacks. These attacks can be carried out by individuals who generally lack advanced technical skills but repurpose off-the-shelf tools and services. Despite their simplicity, these vectors are attractive during wartime as they can be executed at scale and may cause immediately noticeable effects by making digital infrastructure inaccessible or by `painting' sites with unwanted messages and political propaganda.

\lpara{Ethics and Data Availability} We received approval from our institutional ethics committee for data collection and analysis. We only scrape public content and refrain from gathering private data or data that requires authorisation. Our scraper maintains a reasonable pace to avoid burdening websites with unnecessary traffic~\cite{wilson2024identifying}. All analyses are conducted collectively without disclosing individual information to prevent potential harm, aligning with the British Society of Criminology’s Statement on Ethics~\cite{britishethics}. All datasets and scripts are available through a data-sharing agreement with the \ccc.\footnote{Our legal framework: \url{https://cambridgecybercrime.uk/data.html}}

\section{Cyber Operations in Armed Conflicts}
\noindent Similar to Russia and Ukraine, Israel and Palestine have a long-standing antagonistic relationship dating back many years, with Israel possessing high defensive and offensive capabilities. Cyber operations by both state and non-state groups have been reported following the Hamas strike, including destructive attacks such as phishing emails used to deliver data wipers to Israeli organisations~\cite{israelgazaphising}, DDoS attacks targeting Israeli websites that provide information to civilians~\cite{cloudflareisraelhamasddos}, as well as ransomware and website defacements aimed at disrupting infrastructure and spreading political propaganda~\cite{israelgazaorgs}. Iran has conducted hack-and-leak operations and phishing campaigns against Israeli and US entities, while Iranian infrastructure has also been targeted, with disruptions attributed to actors claiming to be retaliating for the conflict~\cite{israelgazatag}. Resembling the IT Army of Ukraine, the IT Army of Palestine was formed in an ad-hoc manner to attract volunteer hacktivists to attack Israeli infrastructure. Though not publicly announced, the group primarily coordinates its activities and advertises its targets through a Telegram channel. However, its scale, operational capacity, and media coverage remain significantly more limited than those of the IT Army of Ukraine.

Iran has also engaged in cyber and influence operations targeting Israel, employing online propaganda and disruptive attacks in support of Hamas~\cite{israelgazairansupport}. Israeli digital billboards were defaced to display Palestinian flags, fabricated war-related news, and political messaging~\cite{israelgazabillboard} as part of a broader disinformation campaign~\cite{israelgazadisinformation}. The pro-Palestinian hacktivist group AnonGhost developed a malicious mobile app that mimicked a legitimate app used by Israeli citizens to receive real-time alerts about incoming airstrikes. This counterfeit application not only misled users with many false warnings but also covertly harvested sensitive personal information and activity logs~\cite{israelgazafakeapp}.

The US has been actively cooperating with Israel on cyber initiatives~\cite{israelgazaushelp}. Although Israel possesses strong cyber warfare capabilities, reports of cyberattacks on Palestinian targets have been limited, largely due to the region's minimal reliance on internet infrastructure. Shortly after the conflict started, Cloudflare observed that over half of all traffic to Palestinian websites were part of HTTP DDoS attacks~\cite{cloudflareisraelhamasddos}. Israel enforced a near-total telecommunications blackout for some time (some lasting an entire week), cutting off internet and telephone connections, which severely impacted critical services including medical operations~\cite{israelgazablackout}. A few days after the Hamas attack, internet accessibility in the Gaza Strip dropped to just 15\% of pre-war levels~\cite{israelgazaconnectivityreduced}. Some pro-Israeli hacktivists have launched cyberattacks, allegedly disrupting Tehran's electrical grid and taking down the Gaza Now news site~\cite{israelgazaelectgriddisabled}.

A comprehensive account of this conflict has been documented~\cite{seljan20247}. Some commentators have suggested that this `cyberwar' has not caused significant harm compared to the physical battlefield, where human lives are at stake. Instead, cyber operations have primarily served as tools for espionage and political propaganda~\cite{israelgazaalookinside}. This conflict once again highlights the role of information operations in modern `hybrid' warfare, where digital tactics complement traditional military actions. It offers an opportunity to analyse attack patterns and compare them to those seen in the Russia-Ukraine conflict, as both are major wars driven by national aspirations but set in different political contexts. While some industry reports have examined the attack landscape~\cite{israelgazatag,cloudflareisraelhamasddos,israelgazairansupport}, academic studies on the role of low-level cybercrime actors in this ongoing conflict -- both quantitative and qualitative -- remain rather limited.

\begin{figure*}[t]
    \centering
    \includegraphics[width=\textwidth]{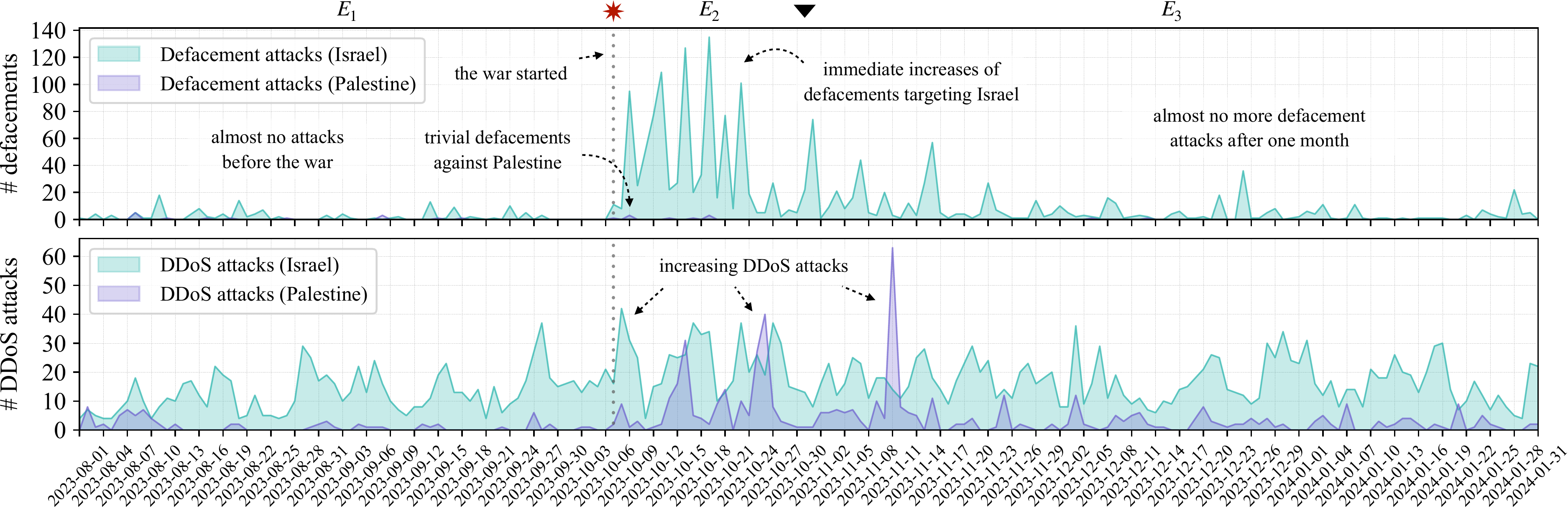}
    \caption{Number of self-reported defacements (top) and DDoS attacks (bottom) on Israel and Palestine over the period. Some attack levels escalated shortly after the war began but then rapidly dropped. The red star marks the Hamas strike.}
    \label{fig:cyberattacks-overview}
\end{figure*}
\begin{table*}[t]
\centering
\normalsize
\caption{The significance of the impact on the number of daily defacements and DDoS attacks on Israel and Palestine.}
\setlength{\tabcolsep}{1.05em}
\begin{tabular}{l|rrrrr}
\toprule
\multirow{2}{*}{Country and type of attacks} & \multicolumn{5}{c}{Statistical tests for the number of attacks per day} \\
\cmidrule{2-6}
& Kruskal-Wallis report & \textbf{$\langle E_1, E_2 \rangle$} & \textbf{$\langle E_1, E_3 \rangle$} & \textbf{$\langle E_2, E_3 \rangle$} & \textbf{$\eta^2$}\\
\midrule
Israel (web defacements) & $H(\CSnewilpsnDefacementsILSignificanceTestDoF) = \CSnewilpsnDefacementsILSignificanceTestStat, \CSnewilpsnDefacementsILSignificanceTestPValue$ & $\CSnewilpsnDefacementsILPostHocPValueONETWO$ & $\CSnewilpsnDefacementsILPostHocPValueONETHREE$ &  $\CSnewilpsnDefacementsILPostHocPValueTWOTHREE$ & \CSnewilpsnDefacementsILEffectSize\\
Palestine (web defacements) & $H(\CSnewilpsnDefacementsPSSignificanceTestDoF) = \CSnewilpsnDefacementsPSSignificanceTestStat, \CSnewilpsnDefacementsPSSignificanceTestPValue$ & \textcolor{lightgray}{$\CSnewilpsnDefacementsPSPostHocPValueONETWO$} & $\CSnewilpsnDefacementsPSPostHocPValueONETHREE$ &  $\CSnewilpsnDefacementsPSPostHocPValueTWOTHREE$ & \textcolor{lightgray}{\CSnewilpsnDefacementsPSEffectSize}\\
\midrule
Israel (DDoS attacks) & $H(\CSnewilpsnDDoSAttacksILSignificanceTestDoF) = \CSnewilpsnDDoSAttacksILSignificanceTestStat, \CSnewilpsnDDoSAttacksILSignificanceTestPValue$ & $\CSnewilpsnDDoSAttacksILPostHocPValueONETWO$ & $\CSnewilpsnDDoSAttacksILPostHocPValueONETHREE$ &  $\CSnewilpsnDDoSAttacksILPostHocPValueTWOTHREE$ & \CSnewilpsnDDoSAttacksILEffectSize\\
Palestine (DDoS attacks) & $H(\CSnewilpsnDDoSAttacksPSSignificanceTestDoF) = \CSnewilpsnDDoSAttacksPSSignificanceTestStat, \CSnewilpsnDDoSAttacksPSSignificanceTestPValue$ & $\CSnewilpsnDDoSAttacksPSPostHocPValueONETWO$ & $\CSnewilpsnDDoSAttacksPSPostHocPValueONETHREE$ &  $\CSnewilpsnDDoSAttacksPSPostHocPValueTWOTHREE$ & \CSnewilpsnDDoSAttacksPSEffectSize\\
\midrule
\end{tabular}
\label{tab:statistical-significance-attacks}
\end{table*}

\section{Methods and Datasets} \label{sec:datasets}
\noindent We analyse the changing landscape of low-level cybercrime activities using quantitative datasets of both cyberattacks and online discussions, spanning from 1 August 2023 to 31 January 2024 -- two months before and four months after the war commenced. As in our prior work~\cite{vu2024getting}, the most substantial changes occurred within six months, making this timeframe sufficient to draw the key narratives. All timestamps are normalised to UTC+0.

\subsection{Datasets}
\noindent We particularly focus on DDoS and website defacement attacks -- two simple and measurable types of cyberattacks that can be launched by low-level cybercrime actors using existing tools and services. These attacks can be executed at scale and may cause immediately visible results during wartime, such as making websites inaccessible or taunting opponents with unwanted political propaganda.

\para{Web Defacements} We analyse \CSnewilpsnIsraelWarTotalDefacements~web defacements within the period, as part of over 350k records shared by the \ccc~\cite{vu2024getting} in a collection of the five most popular archives that defacers use to self-report attacks: \zoneh, \ownzyou, \zonexsec, \haxorid, and \defacerpro. This dataset is self-reported (some actual defacements may be missing); however, combining the most prominent defacement archives provides a reasonably indicative picture. Its reliability and completeness have been comprehensively verified with semi-automatic validation, de-duplication and correction, while victims are identified based on country-code top-level domains (ccTLDs), IP geolocation, and the geolocation of their hosting Autonomous Systems (AS), excluding CDNs~\cite{vu2024getting}.

\para{UDP Amplification DDoS Attacks} We use a DDoS attacks dataset gathered from a honeypot network of several dozen sensors set up worldwide since 2014~\cite{thomas20171000}. This honeypot simulates UDP protocols susceptible to reflective attacks, capturing packets sent by malicious actors but avoid redirecting the magnified traffic to the intended victims. An attack is defined as a flow of at least five packets to a victim, with the victim's country determined by IP and AS geolocation, excluding popular CDNs~\cite{vu2024getting}. With UDP amplification, there is no source information. This dataset covers DDoS attacks caused by many low-level cybercrime actors, particularly booter users, but does not cover TCP-based and direct-path attacks. It has been used to measure booter activity following law enforcement takedown campaigns in 2018~\cite{collier2019booting} and 2022–2023~\cite{vu2025assessing}.

\para{Underground Forum Posts} We analyse war-related discussions on one of the largest hacking forums, \hackforums, as part of the \crimebb dataset~\cite{pastrana2018crimebb}. This forum facilitates cybercrime discussion and trades among low-level actors, some of whom have faced criminal charges~\cite{pastrana2018characterizing}. There are \CSnHFILPSFinalThreads~threads containing at least one post having (case-insensitive) terms `Israel', `Palestine', `Hamas', and `Gaza' over the period: \CSnHFILThreads~related to Israel, \CSnHFPSThreads~related to Palestine, \CSnHFILPSOverlapsThreads~related to both countries, and three 
related to `Hamas' or `Gaza'. We subsequently extracted \CSnILPSHighlyRelevantPosts~posts from \CSnILPSHighlyRelevantThreads~highly relevant threads (with titles directly containing the keywords) and \CSnILPSLowerRelevantPosts~posts having the keywords from the remaining \CSnILPSLowerRelevantThreads~less relevant threads, compiling a total of \CSnILPSAllRelevantPosts~relevant posts made by \CSnILPSAllRelevantUsers~active users.

\para{Telegram Chats} One week after the Hamas strike, the Cyber Army of Palestine was established to support the `digital frontline', mirroring the IT Army of Ukraine~\cite{vu2024getting}. The most tangible outcome is two public Telegram channels, starting on 14 October 2023, to recruit volunteer hacktivists and coordinate attacks against Israeli digital assets. The primary channel is used solely by the admins to spread propaganda while offering training and attack tools, attracting over 13\,000 subscribers (10--15 times less than the IT Army of Ukraine). The secondary channel is used by over 1\,000 participants for social discussions, with announcements forwarded from the primary channel. Successful attacks are frequently showcased in both channels, with victims and domain names often promoted in Arabic; most are Israeli, but sometimes extend to Arab `friends' such as Egypt. Using Telethon and official Telegram APIs, we collect both channels, with 189 admin announcements and 26k messages of over 3\,000 users. Regular expressions are used to extract all targeted domains from the chats.

\subsection{Statistical Tests}
\noindent Similar to our prior work~\cite{vu2024getting}, we test the significance of the impact resulting in different levels of attack counts and discussions by separating the period into three eras; $E_1$: before the Hamas strike, $E_2$: around one immediate following month from 7 October to 31 October 2024, and $E_3$: from 1 November 2023 to 31 January 2024. We then apply the unpaired non-parametric Kruskal-Wallis test (as the data distribution is not normal), with the null hypothesis that there is no significant difference between the three eras. If a difference is found, Dunn’s post-hoc test is used to identify the pairs causing changes. Effect sizes are measured by $\eta^2$, ranging [0, 1]; $0 \leq \eta^2 < 0.01$: no effect; $ 0.01 \leq \eta^2 < 0.06$: small effect; $0.06 \leq \eta^2 < 0.14$: medium effect; and $0.14 \leq \eta^2 \leq 1$: large effect~\cite{miles2001applying}. 

\section{The Evidence of Cyberattacks} \label{sec:cyberattack-evidence}
\noindent This section outlines the evolving landscape of defacement and DDoS attacks targeting both sides. \autoref{fig:cyberattacks-overview} presents the number of attacks per day over the period, while \autoref{tab:statistical-significance-attacks} details the statistical (in)significance of the effect.

\subsection{Website Defacement Attacks} \label{subsec:defacement-attacks}
\begin{figure}[t]
    \centering
    \includegraphics[width=0.4575\textwidth]{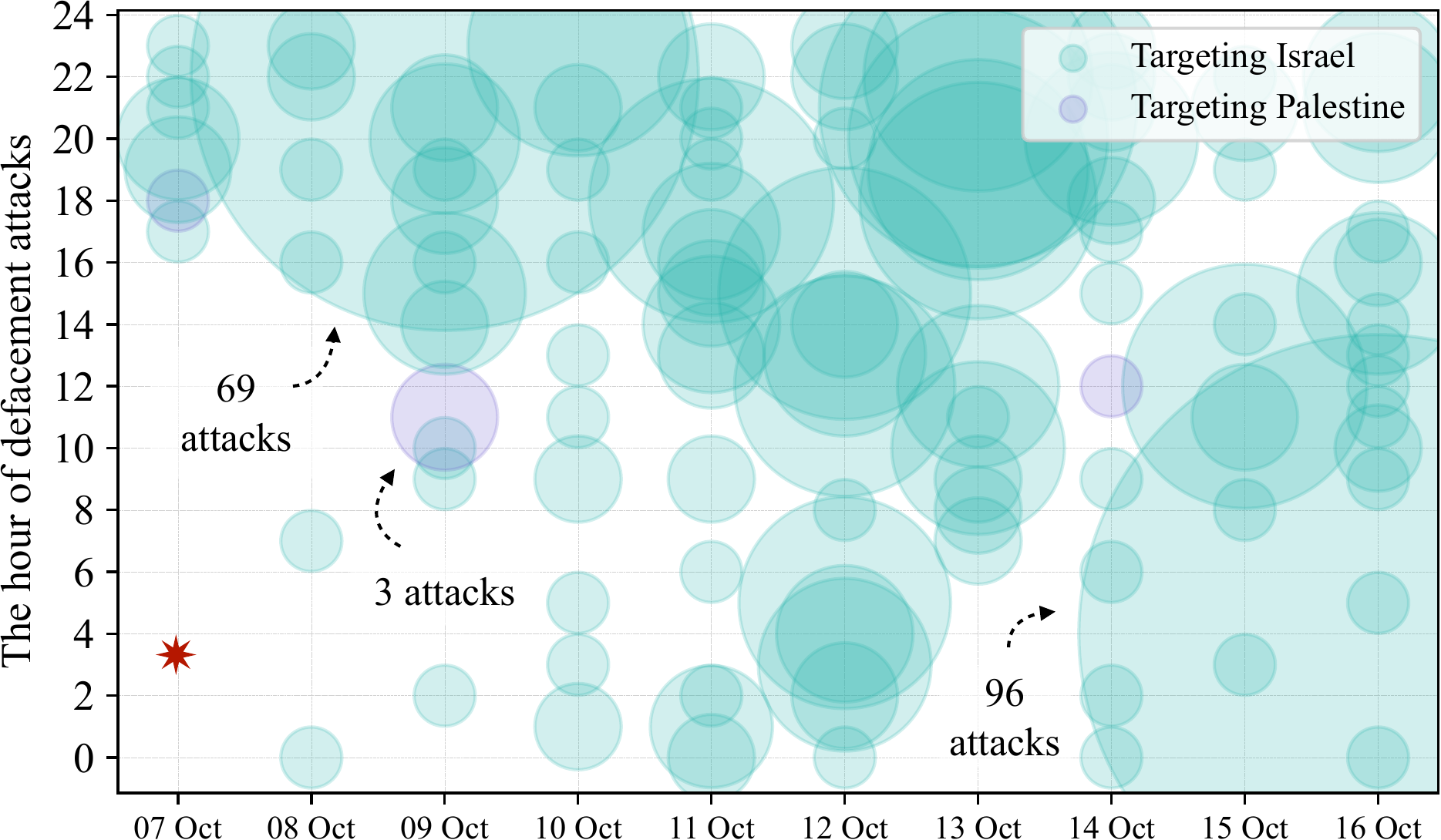}
    \vspace{0.5mm} \caption{The number of website defacement attacks targeting Israel and Palestine, broken down hour by hour. The red star marks the Hamas attack on 7 October 2023.}
    \label{fig:defacements-targeting-israel-hourly}
\end{figure}

\noindent There is evidence of defacement attacks on Israeli websites. While Palestine suffered only \CSnewilpsnIsraelWarDefacementsHittingPalestine~attacks by \CSnewilpsnIsraelWarDefacersHittingPalestine~defacers, we see \CSnewilpsnIsraelWarDefacementsHittingIsrael~attacks on Israel by \CSnewilpsnIsraelWarDefacersHittingIsrael~defacers. There were almost no web defacements targeting Israel in the preceding weeks, with attacks beginning just a few hours after the Hamas attack then escalated quickly (see Figure~\ref{fig:defacements-targeting-israel-hourly}). A small spike occurred on 7 October; the next big one was two days later, following Israel's declaration of war, with around 100 attacks (69 at 10 PM); the peak of nearly 140 attacks was on 19 October. \CSnewilpsnDefacementsILSignificanceTestMethod~test indicates a statistically significant effect on the number of daily defacements against Israel pre-war versus post-war, with a large effect size (\CSnewilpsnDefacementsILEffectSize), suggesting that the conflict is highly associated with the outbreak of attacks on Israel. There is no difference in defacement on Palestine between $E_1$ and $E_2$, primarily due to the small sample size.

\begin{figure}[t]
    \centering
    \includegraphics[width=0.4575\textwidth]{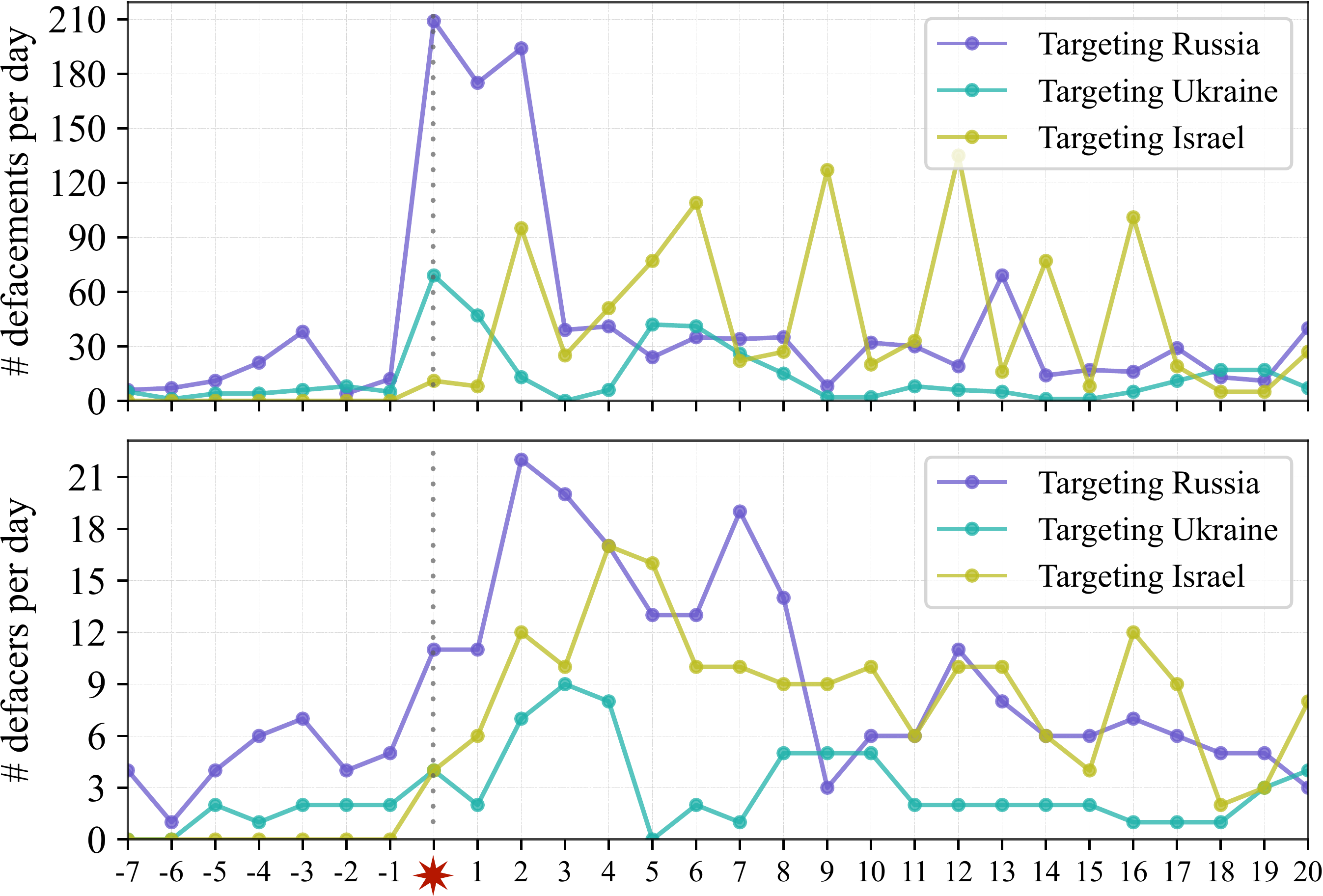}
    \vspace{0.5mm} \caption{Defacements (top) and defacers (bottom) targeting Israel around 7 October 2023, in comparison with those seen with Russia and Ukraine in February 2022~\cite{vu2024getting}.}
    \label{fig:defacements-israel-comparison}
\end{figure}

These outbreaks exhibit some similarities to the cyberattack patterns observed during the Russia-Ukraine conflict, as documented in prior work~\cite{vu2024getting}, see Figure~\ref{fig:defacements-israel-comparison}. The surges in attacks occurred quickly, while the defacer peaks lagged by a few days, presumably as more defacers joined in. Participation then dropped steadily, with only around five attackers still active after three weeks -- following the pattern in the Russia-Ukraine conflict. Offensive activity against Israel (both attacks and attackers) was significantly less than that against Russia but more than that against Ukraine. While attacks targeted both sides in the Russia-Ukraine conflict, the Israel-Gaza war has been mostly one-sided, with no substantial attacks against Palestine so far.

\para{Defacement Motives} Defacers are highly centralised: the 10 most active accounted for \CSnewilpsnIsraelWarTopTenDefacerProps\% of attacks; with one contributing \CSnewilpsnIsraelWarTopOneDefacerProps\%. This aligns with `offender concentration' -- a well-established criminological regularity; for example, just 1\% of repeat offenders were responsible for 57.8\% of repeat defacements~\cite{moneva2022repeat}, and a small number of forum members are involved in the majority of contractual transactions on a cybercrime market~\cite{vu2020turning}. We analyse messages left on defaced pages, considering a political sentiment to be supporting one side if a support/objection is expressed. Signatures without a clear war-related statement are considered self-aggrandisement, and are marked as financially motivated if there are adverts for hacking services e.g., `\textit{contact for shells}'. Among \CSnewilpsnIsraelWarDefacementsTotalAttacksHittingILandPS~defacements on Israel and Palestine, \CSnewilpsnIsraelWarDefacementFunMotives~are self-aggrandisement (\CSnewilpsnIsraelWarDefacementFunMotivesProps\%). Only \CSnewilpsnIsraelWarDefacementProILMotives~supports Israel, but \CSnewilpsnIsraelWarDefacementProPSMotives~defaced Israeli sites in support of Palestine (\CSnewilpsnIsraelWarDefacementProPSMotivesProps\%) with hashtags \#opisrael, \#savegaza, \#freepalestine, and \#savepalestine. That proportion is much higher than what was seen in the Russia-Ukraine conflict, where only around 7\% of attacks explicitly expressed support for either side~\cite{vu2024getting}. We see \CSnewilpsnIsraelWarDefacementFinanceMotives~are financially motivated, and \CSnewilpsnIsraelWarDefacementNationalisticConflictsMotives~expressing warlike sentiment but without a clear supporting side. Beyond defacing Israeli and Palestinian sites, defacers also targeted sites of other countries to express their support or objection to the war. The majority of messages left on the defaced pages similarly show strong support for Palestine, suggesting that the wider opinion is also largely one-sided.

\para{Choice of Targets} There is little evidence of successful defacements against high-profile targets; most victims are corporate, with \CSnewilpsnIsraelWarDomainCatCOIL~attacks (\CSnewilpsnIsraelWarDomainCatCOILProps\%) against businesses under \url{.co.il}, \CSnewilpsnIsraelWarDomainCatORGIL~attacks (\CSnewilpsnIsraelWarDomainCatORGILProps\%) targeting organisations under \url{.org.il}, and \CSnewilpsnIsraelWarDomainCatACIL~attacks (\CSnewilpsnIsraelWarDomainCatACILProps\%) targeting educational websites. The few notable compromised targets include an Israeli housing association, partly exploited on 13 October 2023, a subdomain of the Israel Defense Forces under \url{.idf.il}, and \CSnewilpsnIsraelWarDomainCatACIL~subdomains of the largest public college in Israel. Regarding Palestinian targets, we see only \CSnewilpsnIsraelWarDomainCatPS~\url{.ps} reported victims. The rest are under generic ccTLDs: \url{.com} (\CSnewilpsnIsraelWarDomainCatCOM), \url{.net} (\CSnewilpsnIsraelWarDomainCatNET), \url{.org} (\CSnewilpsnIsraelWarDomainCatORG), \url{.biz} (\CSnewilpsnIsraelWarDomainCatBIZ), \url{.club} (\CSnewilpsnIsraelWarDomainCatCLUB), \url{.art} (\CSnewilpsnIsraelWarDomainCatART), and \CSnewilpsnIsraelWarDomainCatRemaining~unidentifiable IP addresses. As in the Ukraine war, most defacements are strategically unimportant, as defacers usually employ off-the-shelf tools and scan pre-defined ccTLDs to find vulnerable targets for mass defacements.

\subsection{UDP Amplification DDoS Attacks} \label{subsec:ddos-attacks}
\noindent The war appears to be closely linked to a gradual surge in DDoS attacks targeting Palestine, which was almost zero prior to 7 October 2023. The attack counts gradually rose to about 30--40 per day, peaking at over 60 on 11 November 2023. This increase was short-lived, subsiding significantly after one month. There was a slight uptick of 40 attacks targeting Israel per day on 8 October 2023 (one day after the war commenced), however its correlation with the conflict was visually unclear as there have been consistent attacks against Israel for the previous two months. \CSnewilpsnDDoSAttacksILSignificanceTestMethod~test confirms a statistically significant effect on the number of daily DDoS attacks against Israel and Palestine pre-war versus post-war, with medium and large effect sizes (\CSnewilpsnDDoSAttacksILEffectSize~and \CSnewilpsnDDoSAttacksPSEffectSize), suggesting that the conflict is likely associated with these increases.

DDoS attacks targeting Israel seem more enduring than those against Palestine. During the period, the total attacks on Israel captured by the honeypot -- which covers UDP-based but not TCP-based and direct-path attacks -- was five times higher to those hitting Palestine (over 3\,000 vs 600). These volumes, however, have been an order of magnitude less (15--20 times) compared to the scale observed in the Russia-Ukraine conflict, where around 600 attacks per day were recorded using the same dataset~\cite{vu2024getting}. 

\begin{figure*}[t]
    \centering
    \includegraphics[width=\textwidth]{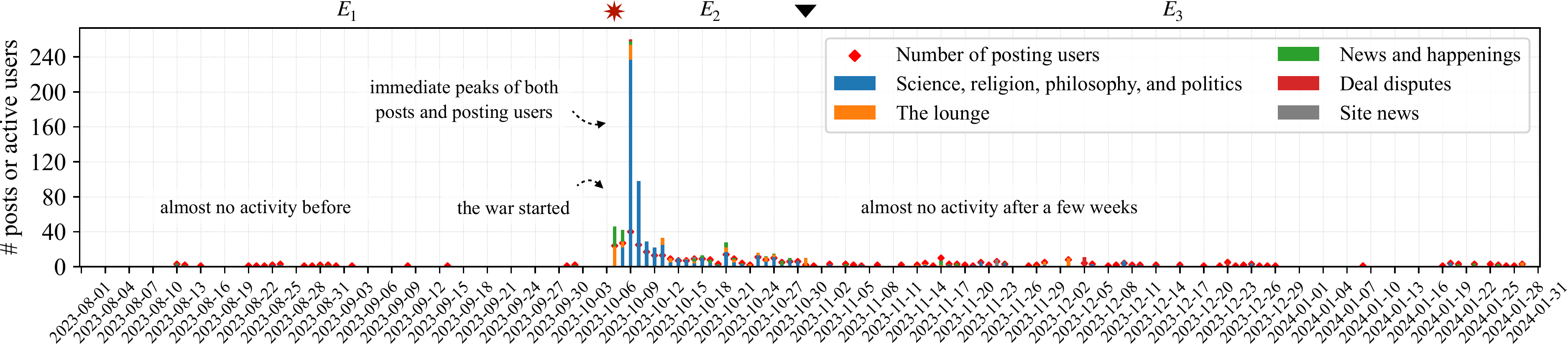}
    \caption{Daily posts and posting users on \hackforums mentioning Israel and/or Palestine (top five subforums).}
    \label{fig:hacking-discussion-about-two-countries}
\end{figure*}
\section{The Hacking Community Reactions} \label{sec:community-reactions}
\para{Discussions on \hackforums} There was an immediate increase in posts related to the conflict on \hackforums from near zero to around 270 per day as the war escalated (see Figure~\ref{fig:hacking-discussion-about-two-countries}). This surge peaked at a higher level but tailed off much faster than those previously seen in the Russia-Ukraine conflict (which peaked at around 140 but lasted for a few weeks~\cite{vu2024getting}), indicating a more short-lived effect. \CSnILPSRelatedPostsSignificanceTestMethod~tests confirm the significance $H(\CSnILPSRelatedPostsSignificanceTestDoF) = \CSnILPSRelatedPostsSignificanceTestStat, \CSnILPSRelatedPostsSignificanceTestPValue$, with a very large effect size~$\eta^2=\CSnILPSRelatedPostsEffectSize$. Pairwise post-hoc tests for $\langle E_1, E_2 \rangle$ and $\langle E_2, E_3 \rangle$ are highly significant ($\CSnILPSRelatedPostsPostHocPValueTWOTHREE$), so is $\langle E_1, E_3 \rangle$ ($\CSnILPSRelatedPostsPostHocPValueONETHREE$). This suggests genuine effects on the hacking community discussions associated with the conflict, with notable shifts particularly from $E_1$ to $E_2$ and $E_2$ to $E_3$.

The number of posting users shows a similar pattern, peaking two days after the Hamas strike. \CSnILPSRelatedUsersSignificanceTestMethod~test reports $H(\CSnILPSRelatedUsersSignificanceTestDoF) = \CSnILPSRelatedUsersSignificanceTestStat, \CSnILPSRelatedUsersSignificanceTestPValue$ with a large effect size~$\eta^2=\CSnILPSRelatedUsersEffectSize$. Pairwise post-hoc tests for $\langle E_1, E_2 \rangle$ and $\langle E_2, E_3 \rangle$ are highly significant ($\CSnILPSRelatedUsersPostHocPValueTWOTHREE$), so is $\langle E_1, E_3 \rangle$ ($\CSnILPSRelatedUsersPostHocPValueONETHREE$). This evidence again suggests significant changes correlated with the war; both the number of posts and users increased sharply, but lagged two days after the Hamas strike -- similar to the defacement activity.

This increased activity is notable as the overall \hackforums activity remained stable at this time. However, this rise is trivial compared to the \hackforums size of around 10k posts per day. We did not see forum users discussing ways to attack either country. Posts are highly centralised: \CSnHFILPSNPostsTopCategoriesProportion\% belong to the top five popular subforums. The biggest subforum, \textit{`science, religion, philosophy, and politics'}, accounts for \CSnHFILPSNPostsTopCategoriesNumberOneProportion\%, and the second biggest, \textit{`the lounge'}, accounts for \CSnHFILPSNPostsTopCategoriesNumberTwoProportion\%. The primary discussion on 7 October was general chats and \textit{`news and happenings'}, but shifted to \textit{`science, religion, philosophy, and politics'} in the following days, while other boards exhibited trivial activity. Interest then declined, dwindling to around 10 posts per day after a week and almost zero after two weeks, suggesting that while users were highly active on 9 October 2023, their engagement waned over time and returned to the previous levels, presumably as they lost interest. This short-lived nature is in line with evidence seen from web defacement and DDoS attacks.

\begin{figure}[t]
    \centering
    \includegraphics[width=0.475\textwidth]{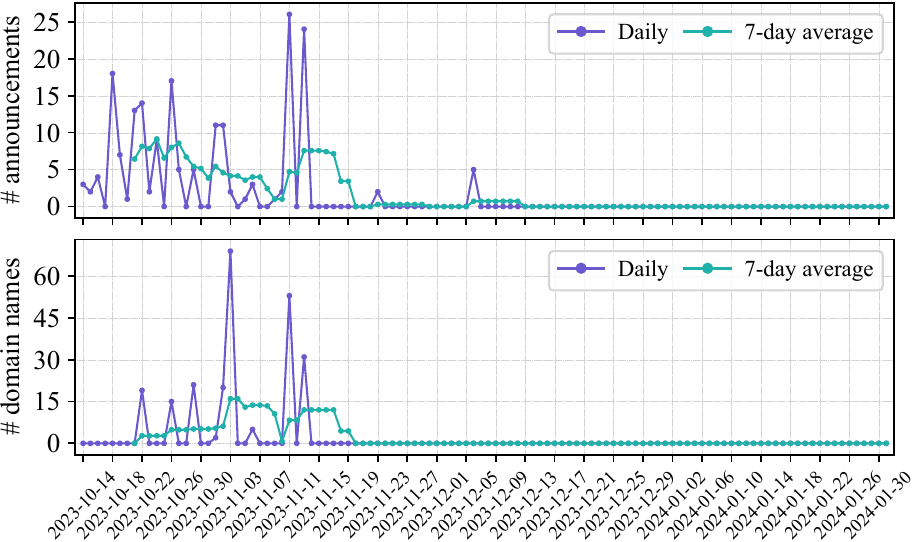}
    \caption{Number of daily announcements and domain names targeted by the Cyber Army of Palestine group.}
    \label{fig:telegram-promoted-targets}
\end{figure}
\begin{figure}[t]
    \centering
    \includegraphics[width=0.475\textwidth]{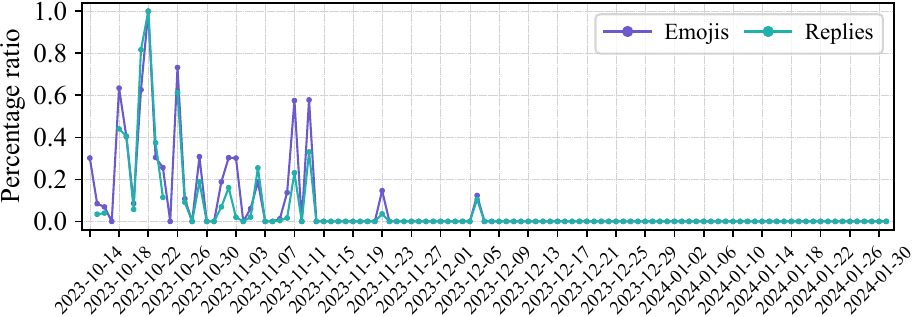}
    \caption{The user engagement (daily) in the Cyber Army of Palestine channel. Values are min–max normalised.}
    \label{fig:telegram-community-reactions}
\end{figure}

\para{The Cyber Army of Palestine} There was a high level of activity in the first month following the channel's inception, peaking at around 25 announcements per day but rapidly dwindling to near zero in subsequent periods (see Figure~\ref{fig:telegram-promoted-targets}). All promoted targets were domains with no associated IP addresses -- unlike the IT Army of Ukraine channel, where our prior work found domains are often associated with IP addresses~\cite{vu2024getting}. Targets were not posted until the second week, peaking at 70 on 3 November 2023. However, as the number of announcements tailed off, the promoted targets also dwindled. The decline in activity extended beyond just the operators; subscribers also exhibited diminishing engagement over time (see Figure~\ref{fig:telegram-community-reactions}). While they actively engaged with announcements during the first month, the number of reactions declined rapidly following the absence of new announcements. We believe the patterns indicate a clear loss of interest from both operators and subscribers. This decline was even more rapid than the declining interest observed with the IT Army of Ukraine~\cite{vu2024getting}, which sustained a prolonged tail of announcements and promoted targets in the two months following the invasion. We do not perform statistical tests on the Telegram chat data, as this channel was established after the conflict, and thus pre-war records are unavailable.

\section{Concluding Remarks} \label{sec:discussions-and-conclusion}
\noindent The role of low-level cybercrime actors in the Israel-Gaza conflict appears to have resembled the timely but short-lived activity in the Russia invasion of Ukraine~\cite{vu2024getting}. These actors were quickly influenced by the war and changed their behaviour, but their interest waned after a few weeks, with both attack levels and war-related discussions gradually dropping. While armed conflicts may be rooted in different ideologies and political contexts, the observed behaviours may also hold in other analogous situations.

Both types of attacks this time were at a much lower intensity than those in the Russia-Ukraine conflict, presumably as Russia and Ukraine have a far longer history of information operations, with substantial offensive capacities~\cite{jaitner2015russian}, and are among the most active cybercrime hubs globally~\cite{lusthaus2020mapping}. Another contrast is that cyberattacks this time were predominantly one-sided: we see many more defacement and DDoS attacks on Israeli than on Palestinian targets, similar to the industry's view on application-layer DDoS attacks~\cite{cloudflareisraelhamasddos}. One significant caveat for this disparity may be the relatively low level and criticality of Internet infrastructure in Palestine, which has far fewer sites than Israel, many of which are hosted on overseas cloud services. The defensive capability, attack surface, and resources are also largely unequal between the two.

There may also be a difference in how people react to the conflict, with multiple reports of rising antisemitic and Islamophobic content online. There is a highly skewed picture of online social media supporting the two sides, with many pro-Palestinian messages spread compared to pro-Israeli content. For instance, hashtags such as \#StandWithPalestine and \#PrayforPalestine attracted billions of views, while \#StandWithIsrael and \#PrayforIsrael generated just a few hundred million on TikTok~\cite{socialmediafavourilpl}. Although this disparity may be partly influenced by complex content recommendation systems, the same picture is also observed in cyberattacks, messages left on defaced websites targeting Israel and Palestine, and also in the broader expressions in defacement attacks targeting other countries.

Our future work will incorporate more recent events. For example, on 1 April 2024, Israel conducted airstrikes on Iranian targets in Syria, killing the highest-ranking Iranian military official. The period from 15 January 2025 to 18 March 2025 (marked by ceasefires) and the surprise attack on the Gaza Strip on 18 March 2025 (ended the ceasefires), will also be measured. Comparisons with other conflicts, such as the long-standing China-Taiwan dispute and those in the Middle East, which present unique socio-political complexities, could offer broader insights into the interplay between conflict, cybercrime, and extremism.

\section*{Acknowledgments}
\noindent We thank Richard Clayton and our colleagues at the Cambridge Cybercrime Centre for their useful feedback. All reviews for this paper are available at \url{https://github.com/wacco-workshop/WACCO/tree/main/WACCO-2025}. This study is supported by the European Research Council (ERC) under the European Union's Horizon 2020 research and innovation programme (grant agreement No 949127).
\bibliographystyle{IEEEtran}
\bibliography{main}

\begin{thebibliography}{10}
\providecommand{\url}[1]{#1}
\csname url@samestyle\endcsname
\providecommand{\newblock}{\relax}
\providecommand{\bibinfo}[2]{#2}
\providecommand{\BIBentrySTDinterwordspacing}{\spaceskip=0pt\relax}
\providecommand{\BIBentryALTinterwordstretchfactor}{4}
\providecommand{\BIBentryALTinterwordspacing}{\spaceskip=\fontdimen2\font plus
\BIBentryALTinterwordstretchfactor\fontdimen3\font minus
  \fontdimen4\font\relax}
\providecommand{\BIBforeignlanguage}[2]{{%
\expandafter\ifx\csname l@#1\endcsname\relax
\typeout{** WARNING: IEEEtran.bst: No hyphenation pattern has been}%
\typeout{** loaded for the language `#1'. Using the pattern for}%
\typeout{** the default language instead.}%
\else
\language=\csname l@#1\endcsname
\fi
#2}}
\providecommand{\BIBdecl}{\relax}
\BIBdecl

\bibitem{hoffman2007conflict}
F.~G. Hoffman,
  \emph{\href{https://www.potomacinstitute.org/images/stories/publications/potomac_hybridwar_0108.pdf}{Conflict
  in the 21st Century: The Rise of Hybrid Wars}}.\hskip 1em plus 0.5em minus
  0.4em\relax Potomac Institute for Policy Studies, 2007,
  \href{https://rb.gy/2u7fpo}{\myurl{https://rb.gy/2u7fpo}}.

\bibitem{jerusalempostddos}
{The Jerusalem Post},
  ``\href{https://www.jpost.com/international/article-768208}{Hackers Hit
  United Hatzalah, Aid Group Treating Wounded Israelis},''
  \href{https://rb.gy/dtl7w1}{\myurl{https://rb.gy/dtl7w1}}, 2023.

\bibitem{oxford2024cyberattacks}
{Oxford Analytica},
  ``\href{https://www.emerald.com/insight/content/doi/10.1108/oxan-db291422/full/html}{Cyberattacks
  on the Middle East Will Continue to Rise},'' \emph{Emerald Expert Briefings},
  2024, \doi{10.1108/OXAN-DB291422}.

\bibitem{israelgazajournalistcasualties}
{Committee to Protect Journalists},
  ``\href{https://cpj.org/2025/02/journalist-casualties-in-the-israel-gaza-conflict/}{Journalist
  Casualties in the Israel-Gaza War},''
  \href{https://rb.gy/hjzftg}{\myurl{https://rb.gy/hjzftg}}, 2025.

\bibitem{vu2024getting}
A.~V. Vu, D.~R. Thomas, B.~Collier, A.~Hutchings, R.~Clayton, and R.~Anderson,
  ``\href{https://anhvvcs.github.io/static/media/vu2024getting.pdf}{Getting
  Bored of Cyberwar: Exploring the Role of Low-Level Cybercrime Actors in the
  {Russia-Ukraine} Conflict},'' in \emph{Proceedings of the ACM World Wide Web
  Conference (WWW)}, 2024, \doi{10.1145/3589334.3645401}.

\bibitem{wilson2024identifying}
L.~Wilson, A.~V. Vu, I.~Pete, and Y.~T. Chua,
  ``\href{https://www.worldscientific.com/doi/10.1142/9781800614079_0015}{Identifying
  and Collecting Public Domain Data for Tracking Cybercrime and Online
  Extremism},'' in \emph{Open-Source Verification in the Age of Google}.\hskip
  1em plus 0.5em minus 0.4em\relax World Scientific, 2024,
  \doi{10.1142/9781800614079_0015}.

\bibitem{britishethics}
\BIBentryALTinterwordspacing
{British Society of Criminology},
  ``\href{http://britsoccrim.org/ethics/}{Statement of Ethics},''
  \href{https://rb.gy/ldwfkl}{\myurl{https://rb.gy/ldwfkl}}, 2015. [Online].
  Available: \url{http://britsoccrim.org/ethics/}
\BIBentrySTDinterwordspacing

\bibitem{israelgazaphising}
{Bleeping Computer},
  ``\href{https://www.bleepingcomputer.com/news/security/eset-partner-breached-to-send-data-wipers-to-israeli-orgs/}{ESET
  Partner Breached to Send Data Wipers to Israeli Orgs},''
  \href{https://rb.gy/0zj5zz}{\myurl{https://rb.gy/0zj5zz}}, 2024.

\bibitem{cloudflareisraelhamasddos}
{Cloudflare},
  ``\href{https://blog.cloudflare.com/cyber-attacks-in-the-israel-hamas-war}{Cyber
  Attacks in the Israel-Hamas War},''
  \href{https://rb.gy/czdpjh}{\myurl{https://rb.gy/czdpjh}}, 2023.

\bibitem{israelgazaorgs}
{The Record},
  ``\href{https://therecord.media/attacks-israeli-orgs-double/}{Attacks on
  Israeli Orgs `More Than Doubled' Since October 7},''
  \href{https://rb.gy/wwmwg2}{\myurl{https://rb.gy/wwmwg2}}, 2024.

\bibitem{israelgazatag}
{Google},
  ``\href{https://blog.google/technology/safety-security/tool-of-first-resort-israel-hamas-war-in-cyber/}{Tool
  of First Resort: Israel-Hamas War in Cyber},''
  \href{https://rb.gy/45b4yz}{\myurl{https://rb.gy/45b4yz}}, 2024.

\bibitem{israelgazairansupport}
{Microsoft},
  ``\href{https://www.microsoft.com/en-us/security/security-insider/intelligence-reports/iran-surges-cyber-enabled-influence-operations-in-support-of-hamas}{Iran
  Surges Cyber-Enabled Influence Operations in Support of Hamas},''
  \href{https://rb.gy/ion04y}{\myurl{https://rb.gy/ion04y}}, 2024.

\bibitem{israelgazabillboard}
{Business Insider},
  ``\href{https://www.businessinsider.com/hackers-infiltrate-israeli-smart-billboards-pro-hamas-messages-2023-10}{Hackers
  Infiltrated Israeli Smart Billboards to Post Pro-Hamas Messages},''
  \href{https://rb.gy/dtfyhd}{\myurl{https://rb.gy/dtfyhd}}, 2023.

\bibitem{israelgazadisinformation}
{Reuters},
  ``\href{https://www.reuters.com/world/disinformation-surge-threatens-fuel-israel-hamas-conflict-2023-10-18/}{Disinformation
  Surge Threatens to Fuel Israel-Hamas Conflict},''
  \href{https://rb.gy/oinh39}{\myurl{https://rb.gy/oinh39}}, 2023.

\bibitem{israelgazafakeapp}
{Cloudflare},
  ``\href{https://blog.cloudflare.com/malicious-redalert-rocket-alerts-application-targets-israeli-phone-calls-sms-and-user-information/}{Malicious
  “RedAlert - Rocket Alerts” Application Targets Israeli Phone Calls, SMS,
  and User Information},''
  \href{https://rb.gy/tu8m25}{\myurl{https://rb.gy/tu8m25}}, 2023.

\bibitem{israelgazaushelp}
{Nextgov/FCW},
  ``\href{https://www.nextgov.com/cybersecurity/2023/10/us-cyber-agencies-very-close-contact-israel-after-unprecedented-hamas-attacks/391156/}{US
  Cyber Agencies in `Very Close Contact' With Israel After Unprecedented Hamas
  Attacks},'' \href{https://rb.gy/xk22yn}{\myurl{https://rb.gy/xk22yn}}, 2023.

\bibitem{israelgazablackout}
{CNN News},
  ``\href{https://edition.cnn.com/2024/01/18/middleeast/gaza-communications-blackout-one-week-israel-hamas-intl/index.html}{Gaza
  Communications Blackout, the Longest of the War, Hits One-Week Mark},''
  \href{https://rb.gy/2axuic}{\myurl{https://rb.gy/2axuic}}, 2024.

\bibitem{israelgazaconnectivityreduced}
{Internet Outage Detection \& Analysis (IODA)},
  ``\href{https://ioda.inetintel.cc.gatech.edu/region/1226?from=1696250808&until=1699400808}{Internet
  Connectivity for Gaza Strip},''
  \href{https://rb.gy/x1un70}{\myurl{https://rb.gy/x1un70}}, 2023.

\bibitem{israelgazaelectgriddisabled}
{Israel National News},
  ``\href{https://www.israelnationalnews.com/news/378811}{`Don't Play With
  Fire': Israeli Hackers Claim to Have Disabled Tehran Electrical Grid},''
  \href{https://rb.gy/ucgprr}{\myurl{https://rb.gy/ucgprr}}, 2023.

\bibitem{seljan20247}
P.~Selj{\'a}n, ``\href{https://doi.org/10.32565/aarms.2024.1.5}{The 7 October
  Hamas Attack: A Preliminary Assessment of the Israeli Intelligence, Military
  and Policy Failures},'' \emph{Academic and Applied Research in Military and
  Public Management Science}, 2024, \doi{10.32565/AARMS.2024.1.5}.

\bibitem{israelgazaalookinside}
{The Conversation},
  ``\href{https://theconversation.com/a-look-inside-the-cyberwar-between-israel-and-hamas-reveals-the-civilian-toll-228847}{A
  Look Inside the Cyberwar Between Israel and Hamas Reveals the Civilian
  Toll},'' \href{https://rb.gy/wxzdwv}{\myurl{https://rb.gy/wxzdwv}}, 2024.

\bibitem{thomas20171000}
D.~R. Thomas, R.~Clayton, and A.~R. Beresford,
  ``\href{https://ieeexplore.ieee.org/document/7945057}{1000 Days of UDP
  Amplification DDoS Attacks},'' in \emph{Proceedings of the APWG Symposium on
  Electronic Crime Research (eCrime)}, 2017, \doi{10.1109/ECRIME.2017.7945057}.

\bibitem{collier2019booting}
B.~Collier, D.~R. Thomas, R.~Clayton, and A.~Hutchings,
  ``\href{https://dl.acm.org/doi/10.1145/3355369.3355592}{Booting the Booters:
  Evaluating the Effects of Police Interventions in the Market for
  Denial-of-Service Attacks},'' in \emph{Proceedings of the ACM Internet
  Measurement Conference (IMC)}, 2019, \doi{10.1145/3355369.3355592}.

\bibitem{vu2025assessing}
A.~V. Vu, B.~Collier, D.~R. Thomas, J.~Kristoff, R.~Clayton, and A.~Hutchings,
  ``\href{https://rb.gy/4nkvvv}{Assessing the Aftermath: the Effects of a
  Global Takedown against DDoS-for-hire Services},'' in \emph{Proceedings of
  the USENIX Security Symposium (USENIX Security)}, 2025,
  \href{https://rb.gy/4nkvvv}{\myurl{https://rb.gy/4nkvvv}}.

\bibitem{pastrana2018crimebb}
S.~Pastrana, D.~R. Thomas, A.~Hutchings, and R.~Clayton,
  ``\href{https://dl.acm.org/doi/10.1145/3178876.3186178}{CrimeBB: Enabling
  Cybercrime Research on Underground Forums at Scale},'' in \emph{Proceedings
  of the ACM World Wide Web Conference (WWW)}, 2018,
  \doi{10.1145/3178876.3186178}.

\bibitem{pastrana2018characterizing}
S.~Pastrana, A.~Hutchings, A.~Caines, and P.~Buttery,
  ``\href{https://link.springer.com/chapter/10.1007/978-3-030-00470-5_10}{Characterizing
  Eve: Analysing Cybercrime Actors in a Large Underground Forum},'' in
  \emph{Proceedings of the International Symposium on Research in Attacks,
  Intrusions, and Defenses (RAID)}, 2018, \doi{10.1007/978-3-030-00470-5_10}.

\bibitem{miles2001applying}
J.~Miles and M.~Shevlin,
  \emph{\href{https://uk.sagepub.com/en-gb/eur/applying-regression-and-correlation/book209139}{Applying
  Regression and Correlation: A Guide for Students and Researchers}}.\hskip 1em
  plus 0.5em minus 0.4em\relax Sage, 2000,
  \href{https://rb.gy/1hi361}{\myurl{https://rb.gy/1hi361}}.

\bibitem{moneva2022repeat}
A.~Moneva, E.~R. Leukfeldt, S.~G. Van De~Weijer, and F.~Mir{\'o}-Llinares,
  ``\href{https://www.sciencedirect.com/science/article/pii/S0747563221003071?via%3Dihub}{Repeat
  Victimization by Website Defacement: An Empirical Test of Premises From an
  Environmental Criminology Perspective},'' \emph{Computers in Human Behavior},
  2022, \doi{10.1016/J.CHB.2021.106984}.

\bibitem{vu2020turning}
A.~V. Vu, J.~Hughes, I.~Pete, B.~Collier, Y.~T. Chua, I.~Shumailov, and
  A.~Hutchings, ``\href{https://dl.acm.org/doi/10.1145/3419394.3423636}{Turning
  Up the Dial: the Evolution of a Cybercrime Market Through Set-Up, Stable, and
  Covid-19 Eras},'' in \emph{Proceedings of the ACM Internet Measurement
  Conference (IMC)}, 2020, \doi{10.1145/3419394.3423636}.

\bibitem{jaitner2015russian}
M.~Jaitner,
  ``\href{https://ccdcoe.org/uploads/2018/10/Ch10_CyberWarinPerspective_Jaitner.pdf}{Russian
  Information Warfare: Lessons From Ukraine},'' in \emph{Cyber War in
  Perspective: Russian Aggression against Ukraine}.\hskip 1em plus 0.5em minus
  0.4em\relax NATO Cooperative Cyber Defence Centre of Excellence (CCDCOE),
  2015, \href{https://rb.gy/ywkisg}{\myurl{https://rb.gy/ywkisg}}.

\bibitem{lusthaus2020mapping}
J.~Lusthaus, M.~Bruce, and N.~Phair,
  ``\href{https://ieeexplore.ieee.org/document/9229673}{Mapping the Geography
  of Cybercrime: A Review of Indices of Digital Offending by Country},'' in
  \emph{Proceedings of the IEEE European Symposium on Security and Privacy
  Workshops (EuroS\&PW)}, 2020, \doi{10.1109/EuroSPW51379.2020.00066}.

\bibitem{socialmediafavourilpl}
{Politico},
  ``\href{https://www.politico.eu/newsletter/digital-bridge/does-social-media-favor-palestine-over-israel/}{Does
  Social Media Favor Palestine Over Israel?}''
  \href{https://rb.gy/z0nwsu}{\myurl{https://rb.gy/z0nwsu}}, 2023.

\end{thebibliography}
\end{document}